\documentclass[]{aastex631}
\usepackage{xcolor}
\usepackage{makecell}

\received{February 12, 2025}
\accepted{July 11, 2025}
\submitjournal{The Planetary Science Journal}

\graphicspath{{./}{figures/}}

\begin{document}
\title{Placing the Near-Earth Object Impact Probability in Context}

\correspondingauthor{C. R. Nugent}
\email{cnugent@olin.edu}

\author[0000-0003-2504-7887]{C. R. Nugent}
\affiliation{Olin College of Engineering\\
1000 Olin Way \\
Needham, MA 02492, USA}


\author[0009-0004-4279-8492]{K. P. Andersen}
\affiliation{
Department of Materials and Production\\
Aalborg University\
Fibigerstr\ae de 16\\
9220 Aalborg, Denmark}

\author[0000-0001-9542-0953]{James M. Bauer}
\affiliation{Department of Astronomy\\
University of Maryland\\
College Park, MD 20742, USA}

\author{C. T. Jensen}
\affiliation{
Department of Materials and Production\\
Aalborg University\
Fibigerstr\ae de 16\\
9220 Aalborg, Denmark}

\author{L. K. Kristiansen}
\affiliation{
Department of Materials and Production\\
Aalborg University\
Fibigerstr\ae de 16\\
9220 Aalborg, Denmark}

\author{C. P. Hansen}
\affiliation{
Department of Materials and Production\\
Aalborg University\
Fibigerstr\ae de 16\\
9220 Aalborg, Denmark}

\author{M. M. Nielsen}
\affiliation{
Department of Materials and Production\\
Aalborg University\
Fibigerstr\ae de 16\\
9220 Aalborg, Denmark}

\author{C. F. Vesterg{\aa}rd}
\affiliation{
Department of Materials and Production\\
Aalborg University\
Fibigerstr\ae de 16\\
9220 Aalborg, Denmark}

\begin{abstract}

Near-Earth objects (NEOs) have the potential to cause extensive damage and loss of life on Earth. Advancements in NEO discovery, trajectory prediction, and deflection technology indicate that an impact could be prevented, with sufficient warning time. We derive an impact frequency of NEOs 140m and larger, using the {\tt\string NEOMOD2}  NEO population model and JPL  {\tt\string Horizons}. We then place that frequency in context with other preventable causes of death; allowing for comparison between a planet-wide event and individual events that cause fatalities such as car crashes and carbon monoxide poisoning.
We find that the chance of a $>140$ m asteroid hitting the Earth is more likely than the chance of an individual being struck by lightning.

\end{abstract}


\section{Introduction} \label{sec:intro}

An asteroid impact is unique among natural disasters; it is the only one that is, theoretically, technologically preventable. The Double Asteroid Redirection Test (DART) Mission demonstrated that a spacecraft could change 
the along-track velocity of a $\sim150$ m diameter asteroid moon by $2.6$ mm s$^{-1}$ \citep{202PNaidu}, corresponding with an orbital period change of
 33 minutes \citep{2023Thomas, 2023Daly}. With sufficient advance notice of years to decades, it is plausible that a similar space mission could change the orbit of a potential impactor of roughly the same size, precluding an impact \citep{NAP12842}. To provide advance notice of an impact, asteroid surveys \citep[e.g.,][]{2015Jedicke} discover new near-Earth objects (NEOs) nightly. Over the last several decades, the number of discovered NEOs has grown.\footnote{Regularly updated data is available at \url{https://cneos.jpl.nasa.gov/stats/totals.html}.} These discoveries have been enabled by legislation such as the American law named ``The George E. Brown, Jr. Near-Earth Object Survey Act'' which mandated NASA discover $90\%$ of all NEOs 140 meters or larger by 2020. NASA has developed a comprehensive planetary defense strategy \citep{2023NASA}.

There is strong evidence that comet and asteroid impacts can be widely destructive. A $\sim10$ km minor planet impact remains the most likely cause of the extinction of the dinosaurs 65 million years ago \citep{1980Alvarez}. This phenomenon is not one of the distant past, as the 1908 Tunguska \citep{1993Chyba}, 1994 Shoemaker–Levy 9 \citep{1995Hammel} and 2013 Chelyabinsk \citep{2013Popova} impacts demonstrate.

When discussing NEO impacts and the destruction they can cause, researchers generally divide the problem into separate domains. 
Some researchers have calculated the frequency of NEO impacts for various size ranges of NEOs. Others have investigated the damaging consequences of impacts. Often, the results of these studies are intended for specialists and are presented without broader context.
\citet[Figure 3]{1994Chapmana} made a notable contribution by placing the probability of being killed due to an NEO impact in relation to other causes of death, such as a tornado or fireworks accident. 

\subsection{Impact Frequencies}

NEO impact probabilities have been estimated with some regularity over the last eighty years. \citet{1994Chapmana} reference an estimate by \citet{1941Watson}, who stated ``the Earth probably goes at least a hundred thousand years between collisions with [minor planets]'' \citep[as quoted by][]{1949Baldwin}. Calculation methodology used in the literature has evolved over time as computational capabilities have increased. For example, \cite{2002Morrison}, following methodology from \citet{1998Harris}, examined all objects discovered as of 3 July 2001 with $H < 18.0$ and perihelion $<1.0$ AU (sample size of 244) and calculated close approaches to determine an impact probability. Those results agreed with those made by \citet{1983Shoemaker} 20 years previously. Recent studies \citep[e.g.,][]{2021Harris, 2024Nesvorny} use more sophisticated numerical integration techniques with a sample of simulated NEOs derived from population models. 

NEO population models, which de-bias survey observations to infer a full picture of the NEO population, can help researchers gain an understanding of NEO origins and evolution over time. They also are used to gauge progress towards discovery mandates. Multiple population models have been created which vary in the methodology specifics and the reference data used; some focus on particular NEO sub-populations of interest \citep{1993Rabinowitz, 2000Bottke, 2000Rabinowitz, 2001DAbramo, 2001Stuart,  2002Bottke,  2011Mainzer, 2012Greenstreet, 2015Harris, 2016Granvik, 2017Tricarico, 2018Granvik, 2021Harris, 2021Heinze, 2023Grav}. This work employs \citet{2024Nesvorny}, which is a update to \citet{2023Nesvorny}, and is based on Catalina Sky Survey observations. It considers asteroids originating from eleven main-belt sources, as well as comets originating from an additional source. NEO population modeling is an active area of investigation. New models were published after the methods described in this paper were completed \citep{2024Nesvornyb, 2025Deienno}, we would not expect our results to be substantially different if these newer models were used.

\subsection{Consequences of an NEO impact}\label{sec:consq}

Quantifying the extent of damage or loss of life from an NEO impact is difficult, due to the wide range of variables at play. This includes (but is not limited to) impactor speed, impactor size, impactor composition (loose regolith, solid rock, metallic), impactor angle, and where the impact occurs (over land or sea, near or far from a populated center). One can divide NEO impacts into two categories; ones that produce only local effects, and ones that produce local and global effects. 

Local effects can include damage from the blast overpressure as the NEO explodes, thermal radiation, and possible tsunamis. To capture the complexity of the possible outcomes, \citet{2017Mathias} used a probabilistic model which sampled from a distribution of asteroid properties to produce millions of simulated impacts across Earth. They considered NEOs $<300$m; these were considered small enough to not produce global effects. They found that in some cases, for impactors 200 m and smaller, an impact in a remote location could occur without affecting a population \citep[Figure 10,][]{2017Mathias}.\citet{2016Rumpf} also considered local effects from NEOs and applied their model over the Earth to determine country-level risk.

Global effects are possible in the case of a high-energy NEO impact. As described in papers by \citet{1994Chapmana} and \citet{1997Toon} these could result in global cooling due to dust, fires, nitrogen oxide generation, the release of sulfur dioxide, multi-continental fires triggered by raining ejecta from the crater site, acid rain, and poisoning from heavy metals from the impacting body. The dust lofting alone has the potential, in some cases, to obscure the sun to the point of stopping photosynthesis, which would then cause a mass extinction. These are in addition to the local effects.

As there is little data on global effects, it is difficult to validate global effect models, and all studies on high energy impacts stress the uncertainties inherent in the results. However, this is the rare case where researchers do not wish for additional data. Toon writes, ``it is to be hoped that no large-scale terrestrial experiments occur to shed light on our theoretical oversights" \citep{1997Toon}. Researchers have made do with the best available proxies, including evidence from historical impacts, volcanic eruptions, and nuclear weapons tests. \citet{2016Reinhardt} created a risk assessment that included local and global effects, though did not include tsunamis. Building on the PAIR model used by \citet{2017Mathias}, \citet{2024Wheeler} used global probabilistic models to evaluate risk. This model is enhanced by the use of an asteroid property inference model \citep{2024Dotson}, and it can be used to assess the risk from a specific NEO (such as the hypothetical NEOs in the Planetary Defense Conference tabletop exercises) as well as global risk. 

An enduring challenge is clearly and accurately conveying the risk from NEOs that results from this modeling. There are many more NEOs of the size that solely cause local effects than NEOs capable of causing global effects upon impact. Taking the average of the risk does not accurately capture this imbalance. The global effects are so severe that the possibility of their occurrence, although less probable than an impact that produces solely local effects, should not be ignored or obscured by averaging. Adding further complexity, as \citet{2002Morrison} explain, estimates often do not take into account social reactions to an impact that could change fatality rates. An effective large-scale evacuation of an impact zone, for example, could save lives. 

We describe a new calculation of impact probability that employs {\tt\string NEOMOD2} and JPL {\tt\string Horizons}. The results are compared to previous calculations. We then compare this probability to the probabilities of other preventable causes of death. Although impact probability calculations have been regularly updated as NEO discoveries have increased, we have not been able to locate a work that contextualizes this frequency in the literature since \citet{1994Chapmana}.

\section{Methods} \label{sec:methods}

The impact frequency calculation was undertaken by a group of six undergraduate students as part of a semester-long project under the guidance of two co-advisors. A simulated set of semi-major axis, eccentricity, and inclination orbital elements as well as absolute magnitude $H$ of $5 \times 10^6$ NEOs was generated using the {\tt\string NEOMOD2} {\tt\string FORTRAN} code \citep{2024Nesvorny}. {\tt\string NEOMOD2}  was chosen because it was a recently published work, and it provided {\tt\string FORTRAN} code to generate the orbital elements. The simulated set of NEOs was restricted such that $10 < H < 22$. This limit is the conventional cutoff that has been used in the literature, as it corresponds to objects  $>140$ m assuming an albedo of 0.14 or 0.15.  $H$ of 22 is a threshold set by the George E. Brown, Jr. Near-Earth Object Survey Act. NEOs $>140$ m cause regional or global devastation upon impact.

Using $H$ as a proxy for size, although common in the literature, has drawbacks. The $H < 22$ limit excludes dark (0.05 albedo) NEOs  $>140$ m. It could also include a subset of objects that are smaller than 140 m but have albedos higher than 0.14. However, there is not evidence that this will bias the results of this work. The NEOMOD2 population, which had associated $H$ values but not sizes, was compared to the NEOMOD3 population, which had $H$ values and diameters \citep[][Figure 15]{2024Nesvornyb}. For an albedo of 0.14 and sizes greater than 0.1 km, there is close agreement in Earth impact flux between the two population models, ruling out a significant excess of large, dark NEOs and small, bright NEOs biasing the results.

 The elements were plotted and compared to the figures in \citet{2024Nesvorny} to ensure accuracy. To complete the set of orbital elements, for each simulated asteroid mean anomaly, longitude of ascending node, and argument of perihelion values were randomly chosen from the valid range of values for these elements.

The trajectories of these simulated asteroids were then calculated, to determine if any would impact Earth over a 150 year interval. This was accomplished by querying the JPL {\tt\string Horizons} ephemeris calculation service \citep{2015Giorgini}\footnote{https://ssd.jpl.nasa.gov/horizons/} for a close approach table via API, providing it with the orbital elements. JPL {\tt\string Horizons} automatically determines the appropriate integration step size and precisely computes trajectories, including close approaches. Its accuracy has been extensively verified by astronomers and spacecraft mission planners.
Submission of API requests and ingesting of responses was handled by code written in {\tt\string Python}. A print interval value  {\tt\string STEP\_SIZE} of 150 years was specified in the API request. To accommodate the large number of requests, the requests were parallelized using the {\tt\string concurrent.futures} {{\tt\string Python} module, and the queries were divided over the six student computers. Total time for the queries was roughly 780 hours. The results of the API queries were collected in a {\tt\string MySQL} database to simplify analysis and enable data verification. An impact was identified when the close approach distance to Earth was less than the radius of the Earth.

To contextualize the NEO impact frequency, literature searches were conducted to locate studies on other causes of death that could be considered preventable. We considered events that affect individuals for this contextualization, instead of other planetary-scale events such as volcanic super-eruptions or large-magnitude earthquakes. Although a comparison to planetary-scale events may be interesting to the small field of experts who are familiar with the Earth's geologic history, it would not be broadly comprehensible to non-specialists. By comparing the NEO impact frequency with familiar events such as influenza illness, we aimed to increase overall understanding of the likelihood of an impact. 

\citet{1994Chapmana} previously placed an asteroid impact in context with other causes of death such as murder, fireworks accidents, and botulism. In that work, they considered the chance of death due to an impact alongside the chance of death due to other factors. This work addresses a slightly different question; we place the chance of an impact occurring anywhere on Earth relative to the chance of other events of concern happening to an individual. This work is therefore intended to provide context to those who wish to know the probability that a $>140$ m impact will occur, anywhere on Earth, in their lifetime. A medium to large Earth-NEO impact would be a remarkable, historic event. It would likely attract media attention, and footage of the impact would likely be recorded and shared worldwide. It would be witnessed by, and would likely emotionally affect, a significant fraction of the human population, even if only a very small fraction of that population was directly affected via loss of life or property. Because of the enormity of an impact event it is worth contextualizing the likelihood that it would happen anywhere on the planet within an average human lifetime.

In the literature search, recent data (2020-2024) was prioritized for higher-frequency events. Studies of lower frequency events often needed a several-decade baseline to accumulate sufficient data to report. Data from peer-reviewed journals and from governmental and intergovernmental organizations was considered. Many studies focused on specific countries. This data was normalized by year and by population; country-wide studies were normalized by the population of the country at the midpoint of the study.

\section{Results}

\subsection{Impact Frequency Calculation}

In the sample of $5 \times 10^6$ simulated NEOs,  three objects were determined to impact the Earth, producing a per-object impact probability of $4 \times 10^{-9}$ yr$^{-1}$. 
Despite the differences in method detail and sample size, this is the same order of magnitude as \citet{2002Morrison}'s per-object probability of $1.68 \times 10^{-9}$ yr$^{-1}$. 

The current number of discovered asteroids thought to be 140 m in diameter or larger (mostly based on absolute magnitude values) is 11,186\footnote{Data from https://cneos.jpl.nasa.gov/stats/totals.html; accessed Jan 18 2025.}. However, to compare our results with \citet{2024Nesvorny}, we use the number of discoveries as of mid-2023, which was 10,502. \citet{2024Nesvorny} places the survey completeness for NEOs brighter than $H=22$ to be roughly $47\%$ \citep[Figure 16]{2024Nesvorny}. \citet{2021Harris} places it a bit higher than $55\%$ \citep[also Figure 16]{2024Nesvorny}, whereas \citet{2023Grav} find $\sim38\%$\footnote{This work notes that their results are consistent with \citet{2021Harris}, citing personal communications and ``recent recalculations in the absolute magnitudes of NEOs.'' This highlights the complexity inherent in these estimates. We include the original \citet{2021Harris} as it is a current peer-reviewed published value.}.  The average of these findings is $46\%$; which we use to infer a total existing population of 
22,800 NEOs in the size range of interest. 

This allows us to estimate an impact frequency of $9.1 \times 10^{-5}$ from our results. Our method is limited by small number statistics. This constraint was imposed by the queries being part of a semester-long project. We consider the case of our results being changed by $\pm1$ impactor to gain a rough sense of uncertainties. Two impactors would have produced an impact frequency of $6.0 \times 10^{-5}$, four an impact frequency of $1.2 \times 10^{-4}$. This range is in line with previous studies. We summarize a subset of results in Table \ref{tab:freq}. 

\begin{deluxetable*}{lCcC}
\tablecaption{Comparison of $>140$m NEO Impact Frequency Calculations over Time}
\tablewidth{1.0\columnwidth}
\tablehead{
\colhead{Study} & \colhead{Magnitude Range}  & \colhead{Size Range} & \colhead{Impact Frequency, per year} 
}
\startdata	
\citet{1941Watson}, quoted in \citet{1949Baldwin}	& $-$	& 	-	 	&  10^{-5} \\
\citet{1958Opik}							& $-$	& $> 130$ m	& 4.6 \times 10^{-5} \\
\citet{1993Rabinowitz}						&  H <22	& $> 140$ m	& \sim10^{-4} \\
\cite{1997Toon}								& $-$	& $> 140$ m	& 5 \times 10^{-4} \\
\citet{2018Granvik}							&  H <22	& $> 140$ m	& \sim10^{-5} \\
\citet{2021Harris}							& H <22	& $>140$ m	& 4 \times 10^{-5} \\
\citet{2024Nesvorny}							& H <22  	& $>140$ m	& 4 \times 10^{-5}  \\
This work 									& H <22 	& $>140$ m 	& $9 \times 10^{-5}$ \\
\enddata
\tablecomments{Size ranges are generally inferred, assuming 0.14 or 0.15 albedo. Albedo assumptions by \citet{1958Opik} are not known.}
\label{tab:freq}
\end{deluxetable*}

One could argue that some fraction of the impact probability is retired, as no currently discovered NEO has a meaningful chance of impact.\footnote{Data from the Center for NEO Studies (CNEOS). https://cneos.jpl.nasa.gov/stats/totals.html; accessed Jan 8 2025.} Using the average of the reported discovery fractions, this would reduce impact probabilities by $38\%$ to $55\%$ \citep{2024Nesvorny, 2023Grav}.

Table \ref{tab:freq} shows that, broadly, recent impact frequency estimates are similar to early estimates such as \citet{1958Opik}, despite the increase of NEO discoveries over time\footnote{The estimates of impact frequencies of $>1$km NEOs has shown more historical variation, this may be partly due to the smaller total numbers of NEOs of this size range. The inferred population of these large objects are $<\sim5\%$ of the population $>140$m.}. This trend has been noted before \citep[e.g.,][]{1994Chapmana, 2002Morrison}. Table \ref{tab:freq} provides further confirmation with the addition of studies from the last 25 years. This implies that observational sampling of the Earth-approaching NEO population has, in a broad sense, accurately reflected the total population. In other words, unless there is a survey discovery bias that has totally escaped detection and persisted for the history of asteroid discovery, undiscovered NEOs are likely to be ``known unknowns'' which have orbital elements in line with known populations, not ``unknown unknowns,'' or a population of NEOs on orbits that are substantially different than discovered NEOs.

A caveat to this conclusion is the population of interstellar objects, or asteroids or comets that originate in an extrasolar system and arrive in or pass through our solar system. Only two are known. The first, `Oumuamua, was discovered in 2017 and the second, Borisov, in 2019. There have been no discoveries since. Although it is likely that these objects are quite rare, we cannot entirely rule out the possibility that they represent a ``unknown unknown'', or a class of object that is systemically missed by surveys.

\subsection{Impact Probability in Context}\label{sec:context}

We compare the NEO impact probability to other technologically-preventable causes of death in Figure \ref{fig:haz}. 
This places an event that would affect the planet (NEO impact) in context with potentially deadly events that can befall \emph{individuals.} This serves the purpose of allowing experts and non-experts to place the probability of an NEO impact within a mental framework of events they may have some familiarity with, such as car crashes and animal attacks. 

Each individual has a unique risk profile, which can vary based on factors such as age, local region, and hobbies. Figure \ref{fig:haz} is intended to provide some familiar deadly events for most audiences; we do not expect all events to be relevant to all readers due to the geographically varied nature of the comparison event studies. An individual could gain context about the NEO impact probability by selectively choosing a subset of events most relevant to them; a different individual, living in a different country, may select a different subset. Individuals might also adjust their own interpretations of Figure 1 based on their behavior. For example, a building inspector who regularly tests and maintains their home carbon monoxide detector may conclude that they have a lower chance than average of experiencing carbon monoxide poisoning. Details on each comparison study, including caveats, are discussed below.  

This figure presents results in terms of frequency over a human lifetime, following research into effective technical communication \citep{17Peters}. The average human lifetime is taken to be 71 years, the current global average.\footnote{World Health Organization. accessed Jan 8 2025. \url{https://www.who.int/data/gho/data/indicators/indicator-details/GHO/life-expectancy-at-birth-(years)}}  The data used to generate Figure \ref{fig:haz} is shown in Table \ref{tab:context}.

\begin{figure}[ht!]
\includegraphics[width=7in,scale=0.5]{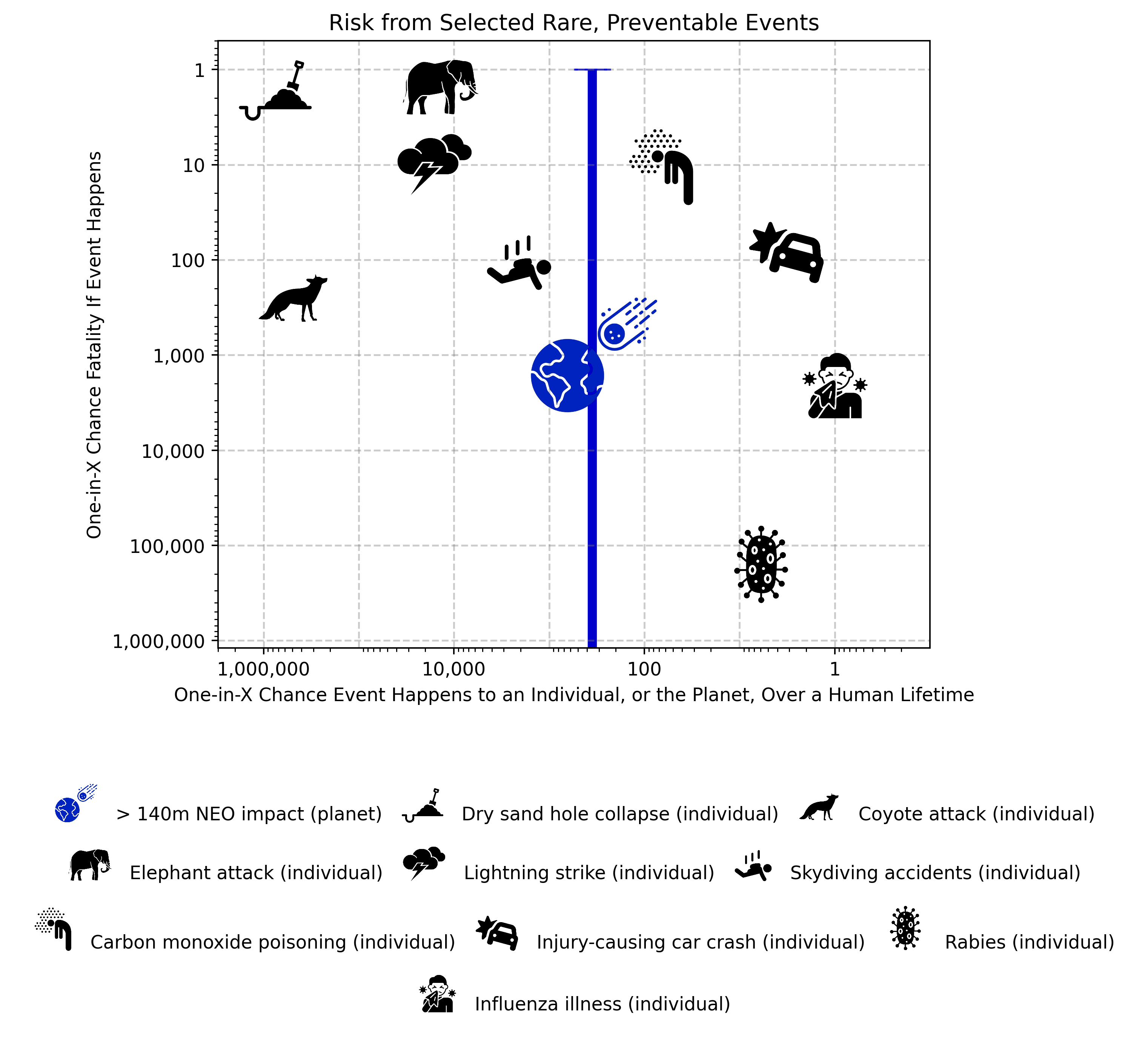}
\caption{The chance of a $>140$m NEO impacting Earth placed in context with the chance that other preventable events may happen to an individual. Individual event frequency and fatality chances are taken from peer reviewed studies on particular regions over specific time intervals; these are broad averages, individual risks vary widely. We expect most readers to be able to identify some events that are relevant to their lives, and some that are not. Readers interested in their own chances are encouraged to consider the events most relevant to their demographics, geographic region, and hobbies. Section \ref{sec:context} has a further discussion of details and caveats. This figure is intended to contextualize the $>140$m NEO impact frequency with events people may be familiar with, such as flu sickness and lightning strikes. The x-axis is the chance of the rare event happening to the planet (NEO impact) or an individual (all other events) over an average human lifetime. The y-axis shows the chance of fatality to an individual if the event occurs. The chance of fatality from a $>140$m impact is dependent on a large range of variables, such as impactor size (140 m to $>10$ km) and impact location (populated city to middle of a large ocean). An impact could result in no fatalities (less likely) some fatalities (more likely) or mass extinction (very unlikely, but possible; see Section \ref{sec:consq}), this is represented by large error bars (blue). The placement of the impact symbol on the y-axis is in the visual center to avoid the pitfalls of averaging discussed in Section \ref{sec:consq}}. 
\label{fig:haz}
\end{figure}

The wide range of outcomes from an NEO impact are represented by large error bars. While a $140-200$m NEO impacting over the ocean could produce no fatalities \citep{2017Mathias,2024Wheeler}, \citet{2017Mathias} (Figure 9) shows that $180-200$m NEOs have small chance of affecting $10^6$ people if an impact was to occur in a highly populated region. The largest NEO impacts have the capability to affect the entire world population \citep{1997Toon, 2024Wheeler}. To avoid the pitfalls of averaging (see Section \ref{sec:consq}) we place the NEO icon in the visual center of the y-axis. 

\begin{deluxetable*}{lcccccl}
\tablecaption{Comparison of Selected Preventable Events}
\tablewidth{1.0\columnwidth}
\tablehead{
\colhead{\parbox{3.5cm}{Event}} &  \colhead{\parbox{2.5cm}{\centering Number Affected Yearly}} & \colhead{\parbox{2.5cm}{\centering Number Killed Yearly}} &
\colhead{\parbox{2.5cm}{\centering Population considered}}  & \colhead{\parbox{2cm}{\centering References}} 
}
\startdata	
\makecell[l]{Dry sand hole collapse (USA)} &  $5.2$ &  $3.1$ &  $2.80\times 10^{8}$ &  \citet{2007Maron} \\
\makecell[l]{Coyote attack (USA)} &  $7.8$ &  $3.00\times 10^{-2}$ &  $2.70\times 10^{8}$ & \citet{2017Baker}\\
\makecell[l]{Elephant attack (Nepal)} &  $2.70\times 10^{1}$ &  $1.80\times 10^{1}$ &  $2.70\times 10^{7}$ & \citet{2016Acharya}\\
\makecell[l]{Lightning strike (USA)} &  $2.77\times 10^{2}$ &  $2.77\times 10^{1}$ &  $3.20\times 10^{8}$ & CDC\\
\makecell[l]{Skydiving accidents (the Netherlands)} &  $1.08\times 10^{2}$ &  $1.0$ &  $1.60\times 10^{7}$ & \citet{2024Damhuis}\\
\makecell[l]{Carbon Monoxide poisoning (Denmark)} &  $1.15\times 10^{3}$ &  $1.05\times 10^{2}$ &  $5.40\times 10^{6}$ & \citet{19Simonsen}\\
\makecell[l]{Injury-causing car crash (Mass, USA)} &  $3.06\times 10^{4}$ &  $3.66\times 10^{2}$ &  $7.10\times 10^{6}$ & Massachusetts DOT\\
\makecell[l]{Rabies (USA)} &  $8.00\times 10^{5}$ &  $5.0$ &  $3.40\times 10^{8}$ & \citet{23Ma}, CDC\\
\makecell[l]{Influenza illness (World)} &  $1.00\times 10^{9}$ &  $4.70\times 10^{5}$ &  $8.10\times 10^{9}$ & WHO\\
\enddata
\tablecomments{Summary of data from papers on preventable fatal events. We calculate the number of people affected by the event each year from the given reference. The number of people killed per year is the number of fatalities due to the event. None of these events are 100\% (or 1:1 chance) fatal. The population considered is the population of the region at the midpoint of time considered by the study; for example, the study on carbon monoxide poisoning examined those events in Denmark between 1995 and 2015. Therefore the population considered is the population of Denmark in 2005. Country and world population estimates are from World Population Prospects, United Nations Population Division\footnote{\url{https://population.un.org/wpp/}}. Massachusetts population value is from the United States Census.\footnote{\url{https://www.census.gov/quickfacts/fact/table/MA/PST045223}}}
\label{tab:context}
\end{deluxetable*}

This comparison contextualizes the NEO impact frequency, providing reference points for the scientific community and enhancing communication with the non-expert public. However, just as NEO impacts are complex and dependent on several variables, so do each of these rare events. The frequency of each of these events is normalized by the relevant country or worldwide population, as appropriate, but this generalization averages over sub-populations that are more vulnerable, such as children. Each of these events, as well as the rationale for why they are considered preventable, are discussed below.

Dry sand hole collapse refers to when a hole is dug, generally at a beach, ``for recreational purposes'' but collapses, trapping someone inside. \citet{2007Maron} writes of their American study, ``The risk of this event is enormously deceptive because of its association with relaxed recreational settings not generally regarded as hazardous.'' It is preventable by not digging large holes in dry sand. The mean age of victims was 12. Older people tend not to dig large holes in the sand.

Deaths from animal attacks and related infections (elephant, coyote, rabies) can be prevented, broadly, by avoiding wild animals, making adjustments to the human-created environment, and medical treatment (such as the rabies vaccine).  
The study on coyote attacks referenced in this work \citep{2017Baker} required a 39 year baseline to establish statistics on this vanishingly rare occurrence in the USA and Canada\footnote{The USA-only results were used in this work.}. They recommend nine programs to prevent coyote-human habituation, including public education. They note that children are much more likely to be the targets of a coyote attack. The study of elephant attacks in Nepal \citep{2016Acharya} has four ``Management recommendations'' to reduce human-elephant encounters, such as ``Restore corridors in critical areas along elephant migratory routes'' so that the elephants can complete their migrations without traveling through populated regions. And although 800,000 Americans seek treatment for rabies following an animal bite yearly, in 2021 only five died; four who did not seek rabies post-exposure prophylaxis treatment and one who did receive treatment but was immunocompromised \citep{23Ma}

The data on lightning strikes was obtained from the American Center for Disease Control (CDC).\footnote{Accessed 27 Jan 2025, \url{https://www.cdc.gov/lightning/data-research/index.html}} Lightning strikes can be prevented by being inside during electrical storms, and avoiding close proximity to plumbing, electrical equipment, windows, and concrete walls and floors reinforced with rebar.\footnote{US Weather Service Lightning Tips, \url{https://www.weather.gov/safety/lightning-tips}} Sub-populations are more vulnerable, for example, lightning is more likely to strike people whose jobs require significant outdoors work.  Skydiving data was taken from a study on the sport in the Netherlands \citep{2024Damhuis}; it is an optional recreational activity. Carbon monoxide poisoning statistics are from Denmark \citep{19Simonsen}, those fatalities can be prevented via the use of working carbon monoxide detectors. 

To the right of the x-axis of the graph are events that are fairly likely to happen to an individual over a lifetime; namely car accidents\footnote{Massachusetts Department of Transportation Crash Data Portal, accessed 21 Jan 2025, \url{https://apps.impact.dot.state.ma.us/cdp/report}} and influenza infections\footnote{World Health Organization, accessed 21 Jan 2025, \url{https://www.who.int/news-room/fact-sheets/detail/influenza-(seasonal)}}. There exist technologies that can reduce the likelihood of death from these events or prevent it,, such as personal protective equipment and advanced safety features. Individuals and societies make trade-offs between the monetary and other costs of this safety equipment, and generally decide to adopt some measures while also assuming some level of risk.

It is impossible to monetarily quantify the full moral and social benefits of saving a single human life, much less quantifying the benefits of preventing the regional to global destruction of an NEO impact. For each of the rare events presented here (with the exception of dry sand hole collapse, which is extremely uncommon), there is often community investment in precautions to reduce risk.  For example, some regional laws require the purchasing of carbon monoxide detectors for each floor of a residence. This results in a cost of roughly 20 to 40 USD over 5-10 years, the lifetime of the detector. Health departments and hospitals stock rabies vaccines. 
 Comparatively, the cost of the NASA-funded DART mission was 324.5 million, or, on average, 1 USD per American. The National Academies report ``Defending Planet Earth'' states, ``The committee considers work on this problem as insurance, with the premiums devoted wholly toward preventing the tragedy.'' \citep{NAP12842}.

\section{Conclusions}
We present a calculation of the impact frequency from NEOs based on an updated model of the NEO population and precision orbital integrations using JPL {\tt\string Horizons}. We show that despite varying methodologies, NEO impact frequency estimates have remained roughly constant over the last 80 years. We place the result in context with other rare, preventable events.

The discovery of a significant fraction of the $ > 140$ m NEO population has enabled detailed studies into the risk from those objects. However, there remains significant work to be done. Asteroid surveys are continuously discovering new NEOs, growing our understanding of the sub-140 m population. This work will eventually improve statistics on the impact frequency of smaller asteroids. 

However, there are sub-populations of Earth-approaching objects that are poorly understood. We have very limited information on interstellar objects from the two known objects in this population. The Vera Rubin Observatory Legacy Survey of Space and Time may shed light on this issue by discovering more interstellar objects \citep{2016Cook, 2023Schwamb}, which may allow for the risk from these objects to be evaluated. 

Despite advances in understanding \citep{1950Oort, 2005Francis, 2017Bauer}, the population of long-period comets is  less well known than the near-Earth asteroid population. Whereas the impact velocity for asteroids is, on average, $20$ km/s, long period comets can have velocities three times as large \citep{1994Chapmana}. As kinetic energy $E = \frac{1}{2} m v^2$, where $m$ is mass and $v$ is velocity, 
comet impact energies can be nine times as large as asteroids of equivalent mass. To enable a precise calculation of the risk from near-Earth comets, there is value in prioritizing the study and discovery of long-period comets, towards creating a comprehensive cometary population model. 

\begin{acknowledgments}
\section{Acknowledgments}
This work was supported by a Fulbright Denmark US Scholar Grant. We thank everyone at the Institut for Materialer og Produktion at Aalborg University for their hospitality. We are grateful to Aalborg University Vice Dean Olav Geil for writing the letter of support that enabled this collaboration and Professor Thomas Tauris for his support, advice, and kindness. We thank T. Spahr for his advice and feedback on this draft. The Olin College library staff obtained access to many of the studies referenced in this work, particular support was provided by M. Anderson. R. Stevenson provided valuable feedback as well. This work was enabled by the JPL {\tt\string Horizons} on-line system. We are grateful to the anonymous referees whose thoughtful comments improved this manuscript.

Figure 1 uses graphical icons (CC BY 3.0) from The Noun Project, 
including 
\href{https://thenounproject.com/browse/icons/term/dig/}{dig} by Adrien Coquet,
\href{https://thenounproject.com/browse/icons/term/fox/}{Fox} by pictohaven,
\href{https://thenounproject.com/browse/icons/term/lightning/}{Lightning} by Dong Gyu Yang 
\href{https://thenounproject.com/browse/icons/term/asteroid/}{asteroid} by Icongeek26,
\href{https://thenounproject.com/browse/icons/term/rabies-virus/}{rabies virus} by Shocho,
\href{https://thenounproject.com/browse/icons/term/flu/}{flu} by Arjuna, 
\href{https://thenounproject.com/browse/icons/term/elephant/}{Elephant} by Harianto,
\href{https://thenounproject.com/browse/icons/term/skydiving/}{Skydiving} by Adrien Coquet,
\href{https://thenounproject.com/browse/icons/term/car-crash/}{car crash} by Tritan Pitaloka,
and
\href{https://thenounproject.com/browse/icons/term/depression/}{depression} by Luis Prado.

\end{acknowledgments}

\software{JPL Horizons  \citep{2015Giorgini}, NEOMOD2 \citep{2024Nesvorny}
          }
\pagebreak

\bibliography{Impact_Bib}{}
\bibliographystyle{psj}

\end{document}